\documentstyle[preprint,tighten,aps,floats,epsfig]{revtex}

\begin{document}
\draft
\title{
The three-loop $\beta$-function of QCD with the clover action. }
\author{A. Bode$^a$ and H. Panagopoulos$^b$\\
}
\address{
$^a$CSIT, Tallahassee, USA\\
{\it email: }{\tt bode@scri.fsu.edu}}
\address{
$^b$Department of Physics, University of Cyprus,\\
P.O.Box 20537, Nicosia CY-1678, Cyprus\\
{\it email: }{\tt haris@ucy.ac.cy}}


\maketitle

\begin{abstract}
We calculate, to 3 loops in perturbation theory, the bare
$\beta$-function of QCD, formulated on the 
lattice with the clover fermionic action.
The dependence of our result on the number of colors $N$, the number of
fermionic flavors $N_f$, as well as the clover parameter $c_{\rm SW}$, is
shown explicitly. 

A direct outcome of our calculation is the two-loop relation between
the bare coupling constant $g_0$ and the one renormalized in the
$\overline{{\rm MS}}$ scheme.

Further, we can immediately derive the three-loop correction 
to the relation between the lattice $\Lambda$-parameter
and $g_0$, which is important in checks of asymptotic scaling.
For typical values of $c_{\rm SW}$, this correction is found to be very
pronounced. 


\end{abstract}

\newpage


\section{Introduction}
\label{introduction}

The clover action for lattice fermions was introduced a number
of years ago~\cite{SW}, as a means of reducing finite lattice spacing
effects, leading to a faster approach to the continuum. It is widely
used nowadays in Monte Carlo simulations of dynamical fermions.

To monitor the onset of the continuum limit, tests of scaling must be
performed on measured quantities. In particular, asymptotic scaling is
governed by the bare $\beta$-function, defined in the standard way:

\begin{equation}
\beta_L(g_0)= -a{dg_0\over da} \mid_{g,\,\mu}, 
\label{lattb}
\end{equation}
($a$ is the lattice spacing, $g$ $(g_0)$ the renormalized (bare)
coupling constant, $\mu$ the renormalization scale). 
In the asymptotic region 
$g_0\rightarrow 0$ one may write $\beta_L$ as:
\begin{equation}
\beta_L(g_0) =
-b_0 \,g^3_0 -b_1 \,g_0^5 - b_2^{L}\,g_0^7 + ...,
\end{equation}
where, in $SU(N)$ gauge theory with $N_f$ fermion species,
\begin{eqnarray}
&&b_0 = {1\over (4\pi)^2} 
\left({11\over 3}N-{2\over 3}N_f\right),\\
&&b_1= {1\over (4\pi)^4} \left[{34\over 3}N^2 - N_f \left(
{13\over 3}N- {1\over N}\right)\right].
\end{eqnarray}

It is well known that the first two coefficients, $b_0$ and $b_1$, are
universal; the 3-loop coefficient $b_2^L$, on the other hand, is
regularization dependent. In the case at hand, $b_2^L$ is thus
expected to depend not only on $N$ and $N_f$, but also on the free
parameter $c_{\rm SW}$ which appears in the clover action (see next Section).

In the present work we calculate $b_2^L$ for arbitrary $N$, $N_f$ and
$c_{\rm SW}$. The analogous calculation for the case of pure gauge theory
without fermions, as well as for the case with Wilson fermions, was
done a few years ago (Refs.~\cite{LWpaper,A-F-P}
and~\cite{C-F-P-V-98}, 
respectively). We follow the general setup and
notation of those publications, to which we refer the reader for
further details.

The $\beta$-function enters directly into the relation defining the
renormalization group invariant parameter $\Lambda_L$ :
\begin{equation}
a\Lambda_L = \exp \left( -{1\over 2b_0g_0^2}\right)
(b_0g_0^2)^{-{b_1/2b_0^2}}
\left[ 1 + q\, g_0^2 + O\left(g_0^4\right)\right], \qquad
q = {b_1^2-b_0b_2^L\over 2b_0^3}.
\label{asympre}
\end{equation}
The ``correction'' factor $q$ is, as we shall see, very pronounced for
typical values of $c_{\rm SW}$ and $g_0$.

A direct outcome of our calculation is the two-loop relation between
the $\overline{\rm MS}$ coupling $\alpha\equiv g^2/(4\pi)$
and $\alpha_0\equiv g_0^2/(4\pi)$ :
\begin{equation}
\alpha = \alpha_0 + d_1(a\mu)\,\alpha_0^2 +
d_2(a\mu)\,\alpha_0^3+O\left( \alpha_0^4\right),
\label{alpha}\end{equation}
This relation is useful in studies
involving running couplings or renormalized quark masses
(see e.g. Refs.~\cite{L-etal-cou2,deDetal,BWW,Gimenez}).
In fact, we will first compute this two-loop relation, which in turn
will allow us to derive $b_2^L$.

In Sec.~\ref{sec2} we describe the method we employed to calculate
$b_2^L$. We present our results for $b_2^L$,
$d_1(a\mu)$ and $d_2(a\mu)$, as functions of $N$, $N_f$ and
$c_{\rm SW}$. Discussion of further technical details and checks of our
calculations is relegated to Sec.~\ref{sec3}.

Given the difficulty of this computation, two different strategies
have been adopted by each of the authors
independently. H.P. performed a standard  perturbation theory
calculation, as described in Section II.\ \   A.B. used the approach of
ref.~\cite{BWW}, in which $d_2(a\mu) - d_2(a\mu)\mid_{c_{\rm SW}=0}$ is
obtained through a two-loop computation of the lattice Schr\"odinger
functional. A detailed comparison of our results is presented at the
end of Section III, showing very good agreeement.

\section{Setup of the Calculation and Results.}
\label{sec2}

Our starting point is the Wilson formulation of the QCD action on the
lattice, with the addition of the clover (SW)~\cite{SW} fermion
term. In standard notation:
\begin{eqnarray}
S_L &=&  {1\over g_0^2} \sum_{x,\,\mu,\,\nu}
{\rm Tr}\left[ 1 - U_{\mu,\,\nu}(x) \right]  +
\sum_{f}\sum_{x} (4r+m_{f,\,0})\bar{\psi}_{f}(x)\psi_f(x) 
\nonumber \\
&-&{1\over 2}\sum_{f}\sum_{x,\,\mu}
\left[ 
\bar{\psi}_{f}(x)\left( r - \gamma_\mu\right)
U_{\mu}(x)\psi_f(x+\hat{\mu})+
\bar{\psi}_f(x+\hat{\mu})\left( r + \gamma_\mu\right)
U_{\mu}(x)^\dagger
\psi_{f}(x)\right]\nonumber \\
&+& {i\over 4}\,c_{\rm SW}\,\sum_{f}\sum_{x,\,\mu,\,\nu} \bar{\psi}_{f}(x)
\sigma_{\mu\nu} {\hat F}_{\mu\nu}(x) \psi_f(x),
\label{latact}
\end{eqnarray}
\begin{equation}
{\rm where:}\qquad {\hat F}_{\mu\nu} \equiv {1\over8}\,
(Q_{\mu\nu} - Q_{\nu\mu}), \qquad 
Q_{\mu\nu} = U_{\mu,\,\nu} + U_{\nu,\,{-}\mu} + U_{{-}\mu,\,{-}\nu} + U_{{-}\nu,\,\mu}
\end{equation}
Here $U_{\mu,\,\nu}(x)$ is the usual product of link variables
$U_{\mu}(x)$ along the perimeter of a plaquette in the $\mu$-$\nu$
directions, originating at $x$;
$r$ is the Wilson parameter; $f$ is a flavor index; $m_{f,\,0}$ are the
bare fermionic masses; $\sigma_{\mu\nu} =
(i/2) [\gamma_\mu,\,\gamma_\nu]$; powers of $a$ have been omitted and
may be directly reinserted by dimensional counting.
The quantity $c_{\rm SW}$ is a free parameter in the present work; it is normally
tuned in a way as to minimize ${\cal O}(a)$ effects.

The procedure we have followed here is analogous
to the one employed in Refs.~\cite{LWpaper,A-F-P,C-F-P-V-98}; we
shall only briefly describe it here, in the interest of a
self-contained exposition.

We set out to compute the relation between the bare lattice coupling $g_0$ 
and the renormalized coupling $g$ defined in the $\overline{\rm MS}$
renormalization scheme: 
\begin{equation}
g_0 = Z_g(g_0,\,a\mu) \, g,
\label{grenorm}
\end{equation}
where $\mu$ indicates the renormalization scale. $Z_g$ is related to
the lattice $\beta$-function $\beta^L(g_0)$, and to its 
$\overline{\rm MS}$-renormalized counterpart $\beta(g)$, through:
\begin{eqnarray}
\beta^L(g_0) &=& -g_0\, a \,{d\over da} \ln
Z_g(g_0,\,a\mu)\mid_{\mu,\,g}\nonumber \\
\beta(g)&=& -g \,\mu\,{d\over d\mu} \ln Z_g(g_0,\,a\mu)\mid_{a,\,g_0}
= -b_0 \,g^3 -b_1 \,g^5 - b_2 \,g^7 + ...
\label{cbf}
\end{eqnarray}
The lattice $\beta$-function, as defined above, is independent of the
renormalized fermionic masses; this follows from the renormalizability of
the theory~\cite{Reisz} combined with dimensional arguments (see
e.g. Ref.~\cite{C-F-P-V-98}). Thus, we will consider fermions with zero
renormalized mass. 

From the above one can derive the relation~\cite{C-F-P-V-98}
\begin{equation}
\beta^{L}(g_0) = 
\left( 1 - g_0^2 {\partial\over \partial g_0^2} \ln Z_g^2 \right)^{-1}
Z_g \,\beta(g_0Z_g^{-1}),
\label{lbcb}
\end{equation}
valid to all orders of perturbation theory.

The one- and two-loop coefficients of $Z_g^2$ have the form:
\begin{equation}
\begin{array}{c}
Z_g(g_0,\,a\mu)^2  = 1 + L_0(a\mu) \,g_0^2 + L_1(a\mu)\, g_0^4 + O(g_0^6)\\[1.0ex]
L_0(x) = 2b_0 \ln x + l_0,\qquad L_1(x) = 2b_1 \ln x + l_1.
\end{array}
\label{zgzg}\end{equation}

The constant $l_0$ is related to the ratio 
of the $\Lambda$ parameters 
associated with the particular
lattice regularization and the
$\overline{\rm MS}$ renormalization scheme:
\begin{equation}
l_{0} = 2b_0\ln \left( \Lambda_L/
\Lambda_{\overline{\rm MS}}\right).
\end{equation}
Its value as a function of $N$, $N_f$, $c_{\rm SW}$, is known (see
e.g. Ref.~\cite{BWW} and references therein) and is presented here
with increased accuracy for the $c_{\rm SW}$-dependent coefficients:
\begin{eqnarray}
&&l_0 ={1\over 8N} -0.1699559991998031(2)\, N + N_f\, l_{01} \label{l0}\\
{\rm where:\ \ } &&l_{01}(r{=}1) = 
 0.006696001(5) -  c_{\rm SW}\, 0.00504671402(1) + {c_{\rm SW}}^2\, 0.02984346720(1)
\nonumber
\end{eqnarray}

We have performed all our calculations for a range of $r$-values. In
what follows we will be presenting only our results for $r=1$, which is the
value used in most Monte Carlo simulations; nevertheless, just for
comparison, we give here $l_{01}$ for $r\ne1$\,:
\begin{eqnarray}
l_{01}(r{=}0.5) &=& \phantom{-}0.020173519(2) - c_{\rm SW} \,0.0144735322(1) +
                    {c_{\rm SW}}^2\, 0.0545958518(1) \nonumber\\
l_{01}(r{=}2.0) &=& -0.00181358(2)\phantom{0} - c_{\rm SW} \,0.0011966285(1) +
                    {c_{\rm SW}}^2 \,0.0119267696(1)
\end{eqnarray}
We see that this range of $r$-values induces significant variations on
$\Lambda_L$\,; the dependence on 
$c_{\rm SW}$ is also quite pronounced, as can be seen from
Eq.~(\ref{l0}), leading to changes in $\Lambda_L$ of up to a factor of 2.

Expanding Eq.~(\ref{lbcb}) in powers of $g_0^2$ leads to a tower of
relations between the coefficients of $\beta^L(g_0)$ and $\beta(g)$\,
; namely, the difference between the $(i{+}1)$-loop coefficients,
$b_i^L - b_i$\,, can be written in terms of $i$-loop contibutions to
$Z_g$. In particular, as is well known, $b_0^L - b_0 = 0$, $b_1^L - b_1
= 0$, and, what will be important below:
\begin{equation}
b_2^L= b_2 -b_1l_{0}+ b_0 l_{1}.
\label{b2lrel}
\end{equation}
Thus, since $b_2$ is known from the continuum~\cite{Tarasovetal}, 
\begin{equation}
b_2 = {1\over (4\pi)^6}
\left[ {2857\over 54}N^3 + N_f \left( -{1709N^2\over 54} + {187\over 36}
+{1\over 4N^2}\right) + 
N_f^2 \left( {56 N \over 27} - {11\over 18N}\right)\right],
\label{b2MS}
\end{equation}
the evaluation of $b_2^{L}$ requires only
a two-loop calculation on the
lattice, of the quantity $l_{1}$.

We will compute $Z_g$ in the background field
gauge~\cite{deWit,Kluberg,Abbott,ellism,LWbg}, where it is simply related to the
background field renormalization constant $Z_A$,
\begin{equation}
Z_A(g_0,\,a\mu)  \, Z_g(g_0,\, a\mu)^2 = 1.
\label{eq:zeqza}
\end{equation}
In the background field formulation the links are written as
\begin{equation}
U_{\mu}(x) = e^{i g_0 Q_{\mu}(x)}\ e^{i a g_0 A_{\mu}(x)},
\end{equation}
where $Q_{\mu}(x)=T^c Q^c_{\mu}(x)$ and
$A_{\mu}(x) = T^c A_{\mu}^c(x)$ are the quantum and
background fields respectively; $T^a$ are Hermitian $su(N)$ generators.
We use the standard gauge-fixing term for this formulation, along with
the corresponding Faddeev-Popov action for ghost fields, and the
contribution from the integration measure (in our notation, these
terms are written explicitly in~\cite{C-F-P-V-98}).

As in Ref.~\cite{LWpaper}, we write the renormalized one-particle irreducible
two-point functions of the background and quantum fields as
\begin{eqnarray}
\Gamma^{AA}_R(p)^{ab}_{\mu\nu} &=
-\delta^{ab}\left( \delta_{\mu\nu}p^2 - p_\mu p_\nu\right)
\left( 1 - \nu_R(p)\right)/g^2,\hfil\qquad\qquad
\nu_R(p) &= \sum_{l=1}^\infty g^{2l} \nu_R^{(l)}(p) , \\
\Gamma^{QQ}_R(p)^{ab}_{\mu\nu} &= 
-\delta^{ab}
\left[ \left( \delta_{\mu\nu}p^2 - p_\mu p_\nu\right)
\left( 1 - \omega_R(p)\right) + \lambda p_\mu p_\nu \right],\hfil\qquad
\omega_R(p) &= \sum_{l=1}^\infty g^{2l} \omega_R^{(l)}(p). 
\end{eqnarray}
Correspondingly on the lattice 
\begin{eqnarray}
\sum_\mu \Gamma^{AA}_L(p)^{ab}_{\mu\mu} &=
-\delta^{ab}3\widehat{p}^2 
\left[ 1 - \nu(p)\right]/g_0^2, \hfil\qquad\qquad
\nu(p) &= \sum_{l=1}^\infty g_0^{2l} \nu^{(l)}(p) ,\\
\sum_\mu \Gamma^{QQ}_L(p)^{ab}_{\mu\mu} &= 
-\delta^{ab}\widehat{p}^2 
\left[ 3\left( 1 - \omega(p)\right) + \lambda_0\right],\hfil\qquad
\omega(p) &= \sum_{l=1}^\infty g_0^{2l} \omega^{(l)}(p). 
\end{eqnarray}

There follows
\begin{equation}
Z_g^2 = Z_A^{-1} = {1 - \nu(p)\over 1 - \nu_R(p)}.
\label{zgzgzg}
\end{equation} 
The bare and renormalized gauge parameters appearing above, $\lambda_0$
and $\lambda$, are related by $\lambda=Z_Q\lambda_0$,
where $Z_Q$ is the renormalization constant of the quantum
field:
\begin{equation}
\left[ 1 - \omega_R(p) \right] = Z_Q 
\left[ 1 - \omega(p) \right].
\end{equation}
Both $\nu(p)$ and $\omega(p)$ depend on $(a\,p)$, $g_0$ and $\lambda_0$,
whereas $\nu_R(p)$ and $\omega_R(p)$ depend on $(p/\mu)$, $g$ and $\lambda$.
One can fix $\lambda_0=1$, and we have done so; this entails computing
$\lambda=Z_Q$, which is only needed here to one loop.

The $\overline{\rm MS}$ renormalized functions 
entering into $Z_g$ to two loops:
$\nu_R^{(1)}(p),\ \omega_R^{(1)}(p)$ and 
$\nu_R^{(2)}(p)\mid_{\lambda=1}$ were calculated and presented in
Ref.~\cite{C-F-P-V-98} (and Ref.~\cite{Ellis} for $N_f=0$).

The lattice two-point functions, $\nu^{(1)}(p),\ \omega^{(1)}(p)$ and 
$\nu^{(2)}(p)$, were the main object of the present work, for
$c_{\rm SW}\ne 0$. We will present here directly our results;
details on our calculations will be provided in the next section.
For easier reference, we also include below results pertaining to pure
gauge theory without fermions (from Refs.~\cite{LWpaper,A-F-P}) and
with fermions at $c_{\rm SW}=0$ (from Ref.~\cite{C-F-P-V-98}).
\begin{eqnarray}
\nu^{(1)}(p)\mid_{\lambda_0=1}\, = &&\nu^{(1)}(p)\mid_{\lambda_0=1,\,N_f=0} + 
N_f \left[{1\over 24\pi^2}\ln (a^2p^2) + k_{1f} \right],
\nonumber\\
\omega^{(1)}(p)\mid_{\lambda_0=1} \,=&& \omega^{(1)}(p)\mid_{\lambda_0=1,\,N_f=0} + 
N_f \left[{1\over 24\pi^2}\ln (a^2p^2) + k_{1f} \right],\\
\nu^{(2)}(p)\mid_{\lambda_0=1} \,= &&\nu^{(2)}(p)\mid_{\lambda_0=1,\,N_f=0} 
+ N_f \left[
{1\over (16\pi^2)^2} \left( 3N - {1\over N}\right)
\ln (a^2p^2) + k_{2f}{1\over N} + k_{3f} N\right].\nonumber
\end{eqnarray}
The $N_f{=}0$ quantities are~\cite{LWpaper}:
\begin{eqnarray}
\nu^{(1)}(p)\mid_{\lambda_0=1,\,N_f=0} &=& -{11N\over 48\pi^2}\ln
(a^2p^2) - {1\over 8N} + N\ 0.217098494366724...\nonumber\\
\omega^{(1)}(p)\mid_{\lambda_0=1,\,N_f=0} &=& -{5N\over 48\pi^2}\ln
(a^2p^2) - {1\over 8N} + N\ 0.137286278290915...\\
\nu^{(2)}(p)\mid_{\lambda_0=1,\,N_f=0} &=& -{N^2\over 32\pi^4}\ln
(a^2p^2) + {3\over 128N^2} - 0.016544619540(4) + N^2\ 0.0074438722(2)
. \nonumber
\end{eqnarray}
For $r=1$ we found
\begin{equation}
\begin{array}{rcclcllcrl}
k_{1f} &=&-&0.013732194(5)&+&c_{\rm SW}&0.00504671402(1)&-&{c_{\rm SW}}^2&0.02984346720(1), \\[0.5ex]
k_{2f} &=& &0.0011877(14) &-&c_{\rm SW}  &0.0001578(3) &-&{c_{\rm SW}}^2&0.0052931(2) \\
   & & & &-&{c_{\rm SW}}^3&0.00050624(3)&-&{c_{\rm SW}}^4&0.00008199(1),\\[0.5ex]
k_{3f} &=&-&0.0013617(16) &-&c_{\rm SW}  &0.0000981(4) &+&{c_{\rm SW}}^2&0.0052440(6) \\
   & & & &+&{c_{\rm SW}}^3&0.00021431(3)&+&{c_{\rm SW}}^4&0.00004382(1).
\end{array}
\label{results}
\end{equation}
The origin and meaning of the errors in the above results
will be explained in the next section. Contributions which vanish in
the continuum limit, $a\to 0$, do not contribute to $Z_g$ and are
neglected as in Refs.~\cite{LWpaper,A-F-P,C-F-P-V-98}.

We have now what is needed to calculate $Z_g$ to two loops
using Eq.~(\ref{zgzgzg}). The resulting two-loop constant $l_1$
is given by
\begin{equation}
l_1 = -{3\over 128 N^2} + 0.018127763034(4) - N^2 \, 0.0079101185(2) 
+ N_f\left[ l_{11}\, {1\over N} + l_{12}\, N\right].
\label{l1}
\end{equation}
\begin{equation}
\begin{array}{rcclcllcrl}
l_{11} &=&-&0.0011967(14)&+&c_{\rm SW} &0.0001578(3) &+&{c_{\rm SW}}^2&0.0052931(2) \\
   & & & &+&{c_{\rm SW}}^3&0.00050624(3)&+&{c_{\rm SW}}^4&0.00008199(1),\\[0.5ex]
l_{12} &=& &0.0009998(16)&+&c_{\rm SW}  &0.0000342(4)&-&{c_{\rm SW}}^2&0.0048660(6) \\
   & & & &-&{c_{\rm SW}}^3&0.00021431(3)&-&{c_{\rm SW}}^4&0.00004382(1).
\end{array}
\label{l11l12}
\end{equation}

At this point, it is a trivial exercise to recover $b_2^L$, for any
value of $N$, $N_f$, $c_{\rm SW}$, by substitution in Eq.~(\ref{b2lrel}).
We list here some particular cases of interest, setting $N=3$:
\begin{equation}
\begin{array}{lcclcrlcrl}
b_2^L(c_{\rm SW}{=}0) &=&-&0.00159983232(13)&+&N_f&0.0000799(4)&-&N_f^2&0.00000605(2) \\[0.5ex]
b_2^L(c_{\rm SW}{=}1) &=&-&0.00159983232(13)&-&N_f&0.0009449(4)&+&N_f^2&0.00006251(2) \\[0.5ex]
b_2^L(N_f{=}3) &=&-&0.0014144(10) &+&{c_{\rm SW}}^{\phantom{3}}  &0.0000654(2)&-&{c_{\rm SW}}^2&0.0024241(4) \\
   & & & &-&{c_{\rm SW}}^3&0.00008108(2)&-&{c_{\rm SW}}^4&0.00001781(1) .
\end{array}
\end{equation}

The correction coefficient $q$ of Eq.~(\ref{asympre}) also follows
immediately. In the particular cases considered above we obtain ($N=3$):
\begin{equation}
\begin{array}{lllllll}
c_{\rm SW}{=}0: &q = 0.18960350(1)&(N_f{=}0),\ \ &q =
                0.2160(1)&(N_f{=}2),\ \ 
                &q = 0.2355(2)&(N_f{=}3)\\[0.5ex]
c_{\rm SW}{=}1: &q = 0.18960350(1)&(N_f{=}0),\ \ &q =
                0.4529(1)&(N_f{=}2),\ \ 
                &q = 0.6138(2)&(N_f{=}3)\\[1.0ex]
\multicolumn{7}{l}{q(N_f{=}3)= 0.2355(2) -c_{\rm SW}\, 0.01007(4) +
                {c_{\rm SW}}^2 \, 0.3731(1) + {c_{\rm SW}}^3 \, 0.01248(1)
                + {c_{\rm SW}}^4 \, 0.002741(1) .}
\end{array}
\end{equation}
Clearly, these values of $q$ bring about a substantial correction to asymptotic
scaling, with a pronounced $c_{\rm SW}$ dependence. 

\medskip
Finally, the coefficients $d_1(a\mu)$ and $d_2(a\mu)$, in the
relation~(\ref{alpha}) between the $\overline{\rm MS}$ coupling
$\alpha$ and $\alpha_0$, can be read off $L_0(x)$ and $L_1(x)$,
defined in Eq.~(\ref{zgzg}):
\begin{equation}
d_1(x)= -4\pi L_0(x), \qquad \qquad d_2(x) = (4\pi)^2\left[ L_0(x)^2-L_1(x)\right]
\label{d1d2}
\end{equation}

\section{The Calculation in Lattice Perturbation Theory}
\label{sec3}

We describe here briefly some technical features and checks of our
calculation of the quantities $\nu^{(1)}(p)$, $\nu^{(2)}(p)$. 
Fermionic contributions to $\omega^{(1)}(p)$ are identical to those in
$\nu^{(1)}(p)$. Many aspects are as in Ref.~\cite{C-F-P-V-98}, and are
discussed there in more detail.

Two diagrams containing fermions contribute to $\nu^{(1)}(p)$, shown
in Figure 1. There are 18 two-loop fermionic diagrams contributing to
$\nu^{(2)}(p)$, as well as two diagrams containing an insertion of the
one-loop fermion mass counterterm; these are shown in Figure 2.

\bigskip
\hskip3.5cm\psfig{figure=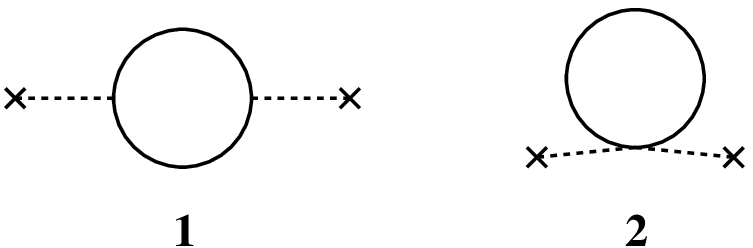,height=2.5truecm}\hskip1.0cm
\nopagebreak\medskip

\noindent
{\small FIG. 1.\ \ Fermion contributions to the one-loop function $\nu^{(1)}(p)$.
Dashed lines ending on a cross represent background gluons.
Solid lines represent fermions.}

\bigskip

The algebra involving lattice quantities was performed using a
symbolic manipulation package in Mathematica, developed by one of the
authors (H.P.) and collaborators over the
past years (see~\cite{C-F-P-V-98} and references therein). In order to
carry out the present work, this package was 
augmented to include vertices of the clover action.
These vertices are more complicated than those of Wilson fermions, as
can be seen by the form of the action~(\ref{latact}); as a result,
the algebraic expression for each diagram is considerably longer in
the present case. On the other hand, all new contributions to
fermionic vertices are accompanied by extra powers of momenta, and
consequently lead to expressions with improved infrared behaviour.

\medskip
The two-loop amplitude $\nu^{(2)}(p)$ can be written as:
\begin{equation}
\nu^{(2)}(p) = \nu^{(2)}(p)|_{c_{\rm SW}{=}0} + \sum_i \nu_i(p)
\end{equation}
(the index $i$ runs over the diagrams shown
in Fig.2). Since two-loop diagrams can have a maximum of 4 fermionic vertices,
$\nu_i(p)$ is a polynomial in $c_{\rm SW}$ of degree up to 4, starting
from ${c_{\rm SW}}^1$.

On general grounds we expect:
\begin{equation}
\widehat{ap}^2 \nu_i(p) = c_{0,\,i} + c_{1,\,i}\, a^2 \sum_\mu {p_\mu^4 \over
p^2} + a^2 p^2 \left\{ c_{2,\,i} \left({\ln a^2 p^2 \over (4 \pi)^2}\right)^2 +
c_{3,\,i}\, {\ln a^2 p^2 \over (4 \pi)^2} + c_{4,\,i} \right \} + 
O((ap)^4) 
\end{equation}
($\widehat{p}^2 = 4 \sum_\mu \sin^2(p_\mu/2)$), where the most general
dependence of $c_{n,\,i}$ on $N,\, N_f,\, c_{\rm SW}$ is:
\begin{equation}
c_{n,\,i} = \sum_{j=1}^4 {c_{\rm SW}}^j \, \left[ c_{n,\,i}^{(-1,\,j)} /N + c_{n,\,i}^{(1,\,j)} N\right] N_f
\end{equation}

\bigskip\bigskip
\hskip0.5cm\psfig{figure=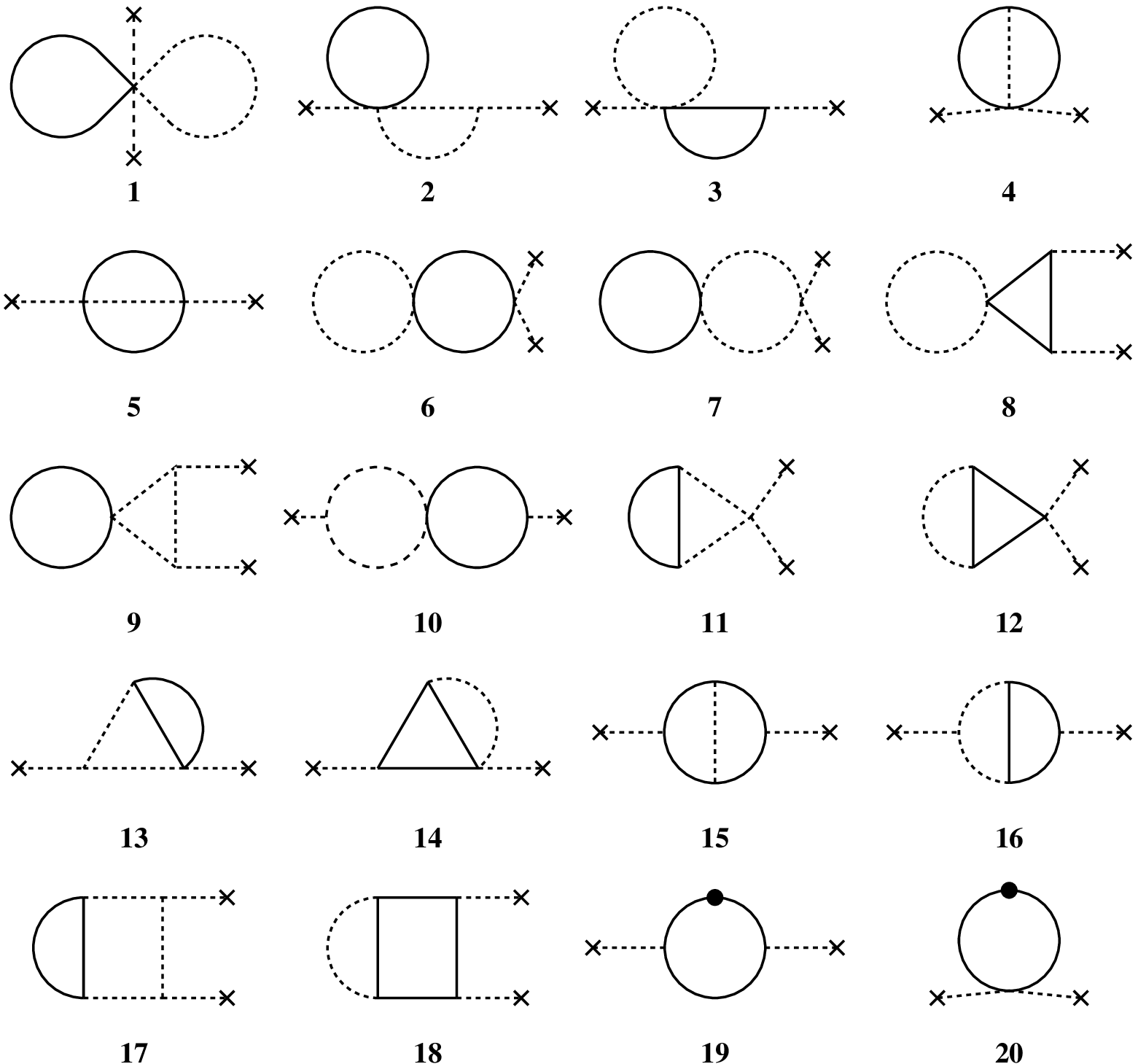,height=13truecm}\hskip1.0cm
\nopagebreak\medskip

\noindent
{\small FIG. 2.\ \ Fermion contributions to the two-loop function $\nu^{(2)}(p)$.
Dashed lines represent gluonic fields;
those ending on a cross stand for background gluons.
Solid lines represent fermions.
The filled circle is a one-loop
fermion mass counterterm.}

\bigskip

As a byproduct of improved infrared convergence, the coefficients
$c_{1,\,i},\ c_{2,\,i}$ vanish for each diagram separately in this case;
all other coefficients are also separately infrared convergent.

To extract the $p$-dependence, we apply a procedure of successive
subtractions~\cite{C-F-P-V-98} on terms containing (sub-)divergences;
our expressions then split into a relatively small part made of
factorized one-loop terms, whose (possibly logarithmic) $p$-dependence
is easily extracted in terms of tabulated lattice integrals, and a
much larger part which can be evaluated by na\"\i ve Taylor
expansion in $ap$. 

Once the $p$-dependence has been made explicit, the resulting
expressions for $c_{n,\,i}$ must be numerically integrated over internal
momenta. For each expression, our package creates highly optimized numerical 
code for integration in momentum space, over finite lattices of size $L\,$.
An extrapolation to infinite size is then performed. This operation is
of course a source of systematic error. In a nutshell, we estimate
this error as follows: First, different
extrapolations are performed using a broad spectrum of functional
forms of the type:
\begin{equation}
\sum_{i,\,j} e_{i,\,j}\, L^{-i}\, (\ln L)^j
\end{equation}
For the $k^{\rm th}$ such extrapolation, a deviation $d_k$ is
calculated applying three different quantitative criteria for quality of
fit. Finally, these 
deviations are used to assign weights $d_k^{-2}/(\sum_k d_k^{-2})$ to
each extrapolation, which are then combined to produce a final value together with
the error estimate. We have checked the validity of these estimates in
cases where the exact answer was known, finding the
estimate to be always correct. 

The improved infrared behaviour of each diagram led to a rather stable
extrapolation; indeed, for the accuracy reported in this work,
lattices up to $L\le 28$ were sufficient in all but a few cases.

The results for each diagram are presented in Tables I through VI. Diagrams
not appearing in these Tables give vanishing contributions.

For the one-loop amplitude $\widehat{ap}^2\,\nu^{(1)}(p)$, the individual
contributions from the corresponding two diagrams in Figure 1 are:
\begin{equation}
\nu_1(p) = c_{\rm SW}\, 0.00504671402(1) - {c_{\rm SW}}^2\,
0.02984346720(1),  \qquad \nu_2(p) = 0
\end{equation}

\medskip
There exist several constraints on the coefficients $c_{n,\,i}$,
which we have used as verification both on the algebraic expressions
and on the numerical results:

\begin{itemize}
\item[$\rhd$]
{\underline {$c_{0,\,i}$}}: Gauge invariance requires 
$\sum_i c_{0,\,i} = 0.$
This relation amounts to 4 independent nontrivial checks, for the
coefficients of $c_{\rm SW}\, N$, ${c_{\rm SW}}^2\, N$, $c_{\rm SW}/ N$,
${c_{\rm SW}}^2/ N$. Our algebraic expressions pass these checks. Our
numerical results also give zero, well within the systematic errors; this
effectively controls the consistency of our error estimation.

\item[$\rhd$]
{\underline {$c_{3,\,i}$}}: The total contribution for single logarithms
must give zero: $\sum_i c_{3,\,i} = 0$.
Again, this relation amounts to 5 independent nontrivial checks, which
were performed both algebraically and numerically.

\item[$\rhd$]
There are several other relations among the coefficients appearing
in individual diagrams (e.g. $c^{(1,\,1)}_{3,\,10}= 4\,c^{(1,\,1)}_{3,\,13},
\ c^{(1,\,2)}_{3,\,17}= -5 \,c^{(1,\,2)}_{3,\,13}$, etc.); these can be
understood in terms of the vertices entering these diagrams, and have
been used as further consistency controls.

\end{itemize}

The most stringent check is provided by Ref.~\cite{BWW}, in
which the Schr\"odinger functional in lattice QCD is computed to two
loops, using the clover action with $c_{\rm SW}=0,\ 1$. Eq. (5.6)
of that reference\footnote{In the notation of Ref.~\cite{BWW}:
$c_{\rm SW} = c_{\rm SW}^{(0)} + c_{\rm SW}^{(1)}\, g_0^2 + \ldots$,\\ 
\phantom{In the notation of Ref. [7]:}$d_2(a\mu) = d_1(a\mu)^2
+ \sum_{r{=}0}^1 N_f^r \{-32 \pi^2 b_{1r} 
\ln(a\mu) + d_{2r}\}$} 
reads:  
$d_{21}(N{=}3,\,c_{\rm SW}^{(0)}{=}1) = 1.685(9) - 8.6286(2) \, c_{\rm SW}^{(1)}$. 
Calculating the same quantity from our results, we deduce: 
$1.6828(8) - 8.62843775(1) \, c_{\rm SW}^{(1)}$. 
Both sets of numbers are clearly in very good agreement.

Since the dependence of $d_2$ on $c_{\rm SW}$ is polynomial of degree
4, all coefficients of the polynomial (except for the 0th degree,
which is gotten from~\cite{C-F-P-V-98}) can be checked by repeating
the calculations of Ref.~\cite{BWW} for a few more values of
$c_{SW}$. We have done so for all values of $c_{\rm SW}$ listed in
Table VII; the results obtained in this way have a larger error than
those stemming from Eqs. (\ref{d1d2}, \ref{zgzg}, \ref{l1},
\ref{l11l12}), but they provide a very solid and 
thorough confirmation of our main result, Eq.(\ref{l11l12}).

\medskip
The entire procedure employed in this paper can be applied without
modifications to a number of other cases of interest, among them the
higher order renormalization of ${\cal O}(a)$-improved operators.
We hope to study these cases in a future work.

\bigskip\bigskip
\noindent
{\bf Acknowledgements:} H.P. would like to thank P. Weisz for many
helpful discussions and suggestions.

\bigskip

\begin{table}[h]
\begin{minipage}{3cm}
\hfill
\end{minipage}
\begin{minipage}{10cm}
\caption{Coefficients $c^{(-1,\,j)}_{0,\,i}$. $r=1$. 
\label{tab1}}
\begin{tabular}{cr@{}lr@{}l}
\multicolumn{1}{c}{$i$}&
\multicolumn{2}{c}{$c^{(-1,\,1)}_{0,\,i}$}&
\multicolumn{2}{c}{$c^{(-1,\,2)}_{0,\,i}$}\\
\tableline \hline
4  &   0&.000040200(3) &   0&\\
12 &   0&.00114029(4)  &   0&.00114886(4)\\
14 &$-$0&.00008041(1)  &   0&\\
15 &$-$0&.00013096(2)  &$-$0&.00007625(1)\\
18 &$-$0&.0009691(2)   &$-$0&.0010726(1)\\
19 &   0&.002862565(1) &   0&.001191276(1)\\
20 &$-$0&.0028625645(1)&$-$0&.0011912762(1)\\
\end{tabular}
\end{minipage}

\bigskip

\begin{minipage}{3cm}
\hfill
\end{minipage}
\begin{minipage}{10cm}
\caption{Coefficients $c^{(1,\,j)}_{0,\,i}$. $r=1$. 
\label{tab2}}
\begin{tabular}{cr@{}lr@{}l}
\multicolumn{1}{c}{$i$}&
\multicolumn{2}{c}{$c^{(1,\,1)}_{0,\,i}$}&
\multicolumn{2}{c}{$c^{(1,\,2)}_{0,\,i}$}\\
\tableline \hline
4  &$-$0&.00016080(1) &   0&.0071556(1)\\
5  &   0&.000040202(2)&$-$0&.0077678(3)\\
11 &$-$0&.00021503(4) &   0&.0023875(1)\\
12 &$-$0&.00114029(4) &$-$0&.00114886(4)\\
13 &$-$0&.0000575(1)  &   0&.000956(1)\\
14 &   0&.00029874(1) &   0&.00026811(2)\\
16 &   0&.00004358(1) &$-$0&.00011562(2)\\
17 &   0&.0002220(2)  &$-$0&.002807(1)\\
18 &   0&.0009691(2)  &   0&.0010726(1)\\
19 &$-$0&.002862565(1)&$-$0&.001191276(1)\\
20 &   0&.0028625645(1)&  0&.0011912762(1)\\
\end{tabular}
\end{minipage}

\bigskip

\caption{Coefficients $c^{(-1,\,j)}_{3,\,i}$. $r=1$. 
\label{tab3}}
\begin{tabular}{cr@{}lr@{}lr@{}l}
\multicolumn{1}{c}{$i$}&
\multicolumn{2}{c}{$c^{(-1,\,1)}_{3,\,i}$}&
\multicolumn{2}{c}{$c^{(-1,\,2)}_{3,\,i}$}&
\multicolumn{2}{c}{$c^{(-1,\,3)}_{3,\,i}$}\\
\tableline \hline
8  &$-$0&.30986678046212 &   0&&0&\\
14 &$-$0&.01449434463(1) &   0&&0&\\
15 &   0&.00500026042(1) &$-$0&.00589886486(1)&   0&\\
18 &$-$0&.0933193205(1)  &   0&.0566733951(1) &   0&.03619153756(1)\\
19 &   0&.41268018519(1) &$-$0&.05077453020(1)&$-$0&.03619153756(1)\\
\end{tabular}
\end{table}

\begin{table}
\caption{Coefficients $c^{(1,\,j)}_{3,\,i}$. $r=1$. 
\label{tab4}}
\begin{tabular}{cr@{}lr@{}lr@{}l}
\multicolumn{1}{c}{$i$}&
\multicolumn{2}{c}{$c^{(1,\,1)}_{3,\,i}$}&
\multicolumn{2}{c}{$c^{(1,\,2)}_{3,\,i}$}&
\multicolumn{2}{c}{$c^{(1,\,3)}_{3,\,i}$}\\
\tableline \hline
8  &   0&.30986678046212&   0&&0&\\
10 &   0&.01009342804(1)&$-$0&.11937386878(1)&   0&\\
13 &   0&.00252335701(1)&$-$0&.02984346720(1)&   0&\\
14 &$-$0&.00500026042(1)&   0&.01052896544(1)&   0&\\
16 &   0&.0271111297(1) &$-$0&.00463010057(1)&   0&\\
17 &$-$0&.02523357012(1)&   0&.14921733598(1)&   0&\\
18 &   0&.0933193205(1) &$-$0&.0566733951(1) &$-$0&.03619153756(1)\\
19 &$-$0&.41268018519(1)&   0&.05077453020(1)&   0&.03619153756(1)\\
\end{tabular}

\bigskip\bigskip

\caption{Coefficients $c^{(-1,\,j)}_{4,\,i}$. $r=1$. 
\label{tab5}}
\begin{tabular}{cr@{}lr@{}lr@{}lr@{}l}
\multicolumn{1}{c}{$i$}&
\multicolumn{2}{c}{$c^{(-1,\,1)}_{4,\,i}$}&
\multicolumn{2}{c}{$c^{(-1,\,2)}_{4,\,i}$}&
\multicolumn{2}{c}{$c^{(-1,\,3)}_{4,\,i}$}&
\multicolumn{2}{c}{$c^{(-1,\,4)}_{4,\,i}$}\\
\tableline \hline
3  &   0&.00082631538(1)&$-$0&.007460866799(1)& 0&&0&\\
4  &   0&.000020100(1)  &$-$0&.00119261(2)  &   0&&0&\\
8  &   0&.003071859(3)  &   0&.00121700384(1)&  0&&0&\\
11 &$-$0&.000020100(1)  &   0&.00059631(2)  &   0&&0&\\
14 &   0&.00018663(2)   &   0&.000005901(1) &   0&&0&\\
15 &   0&.0000931(2)    &   0&.0001160(1)   &$-$0&.00000655(2)  &$-$0&.000038170(1)\\
18 &   0&.0004905(4)    &$-$0&.0002038(3)   &$-$0&.00059685(2)  &   0&.00008405(1)\\
19 &$-$0&.004826287(1)  &   0&.0016290054(5)&   0&.0000971639(1)&$-$0&.0001278777(1)\\
\end{tabular}

\bigskip\bigskip

\caption{Coefficients $c^{(1,\,j)}_{4,\,i}$. $r=1$. 
\label{tab6}}
\begin{tabular}{cr@{}lr@{}lr@{}lr@{}l}
\multicolumn{1}{c}{$i$}&
\multicolumn{2}{c}{$c^{(1,\,1)}_{4,\,i}$}&
\multicolumn{2}{c}{$c^{(1,\,2)}_{4,\,i}$}&
\multicolumn{2}{c}{$c^{(1,\,3)}_{4,\,i}$}&
\multicolumn{2}{c}{$c^{(1,\,4)}_{4,\,i}$}\\
\tableline \hline
3  &$-$0&.00093961433(1) &   0&.008800841023(1)& 0&&0&\\
4  &$-$0&.0000050250(3)  &   0&.000298152(4) &   0&&0&\\
5  &   0&.0000004358(3)  &$-$0&.0001706(1)   &   0&&0&\\
8  &$-$0&.003071859(3)   &$-$0&.00121700384(1)&  0&&0&\\
10 &$-$0&.000169546987(1)&   0&.002005213655(1)& 0&&0&\\
11 &   0&.00003775(1)    &$-$0&.00074560(4)  &   0&&0&\\
13 &   0&.00004011(2)    &$-$0&.0005579(2)   &   0&&0&\\
14 &$-$0&.0002223(1)     &$-$0&.0001602(1)   &$-$0&.00001405(1)&0&\\
16 &$-$0&.0001813(4)     &$-$0&.0000980(5)   &$-$0&.00027132(2)&0&\\
17 &   0&.0000775(3)     &$-$0&.001486(1)    &   0&&0&\\
18 &$-$0&.0004905(4)     &   0&.0002038(3)   &   0&.00059685(2)&$-$0&.00008405(1)\\
19 &   0&.004826287(1)   &$-$0&.0016290054(5)&$-$0&.0000971639(1)&0&.0001278777(1)\\
\end{tabular}
\end{table}

\bigskip\bigskip

\begin{table}[ht]
\caption{Comparison of our results for $(d_{21}\mid_{c_{\rm SW}^{(0)}}
- d_{21}\mid_{c_{\rm SW}^{(0)}=0})/(16\pi^2)$
coming from standard perturbation theory (PT)
(Eqs. (\ref{d1d2}, \ref{zgzg}, \ref{l1}, \ref{l11l12})), to those we
obtain using the Schr\"odinger functional (SF) methods of Ref. [7].
\ \ ($N=3,\ c_{\rm
SW}^{(1)}=0$) 
}
\medskip
\begin{minipage}{3cm}
\hfill
\end{minipage}
\begin{minipage}{10cm}
\begin{tabular}{r@{}lr@{}lr@{}l}
\multicolumn{2}{c}{$c_{\rm SW}^{(0)}$}&
\multicolumn{2}{c}{PT}&
\multicolumn{2}{c}{SF}\\
\tableline \hline
 -4&.0&\quad0&.202270(29)  &\quad0&.2029(21)   \\
 -2&.0&     0&.049518(10)  &     0&.04968(53)  \\
 -1&.0&     0&.012619(7)   &     0&.013041(60) \\
 -0&.5&     0&.003233(2)   &     0&.003258(96) \\
  0&.5&     0&.003197(2)   &     0&.003191(18) \\
  1&.0&     0&.013257(7)   &     0&.01327(4)   \\
  2&.0&     0&.056484(10)  &     0&.05659(45)  \\
  4&.0&     0&.261725(29)  &     0&.2627(33)   \\
\end{tabular}
\end{minipage}
\end{table}



\begin{references}

\bibitem{SW} B. Sheikholeslami and R. Wohlert, Nucl. Phys. {\bf B259}
(1985) 572.

\bibitem{LWpaper} M. L\"uscher and P. Weisz, Nucl. Phys. {\bf B452}
(1995) 234.

\bibitem{A-F-P} B. All\'es, A. Feo and H. Panagopoulos,
Nucl. Phys. {\bf B491} (1997) 498.

\bibitem{C-F-P-V-98}
C.~Christou, A.~Feo, H.~Panagopoulos and E.~Vicari,
Nucl. Phys. {\bf B525} (1998) 387; erratum-ibid. {\bf B608} (2001) 479.

\bibitem{L-etal-cou2} M. L\"uscher,
R.~Sommer, P. Weisz and U.~Wolff,  Nucl. Phys. {\bf B389}
(1993) 247; Nucl. Phys. {\bf B413}
(1994) 481.

\bibitem{deDetal} G.~de Divitiis, R.~Frezzotti, 
M.~Guagnelli and R.~Petronzio, 
Nucl. Phys. {\bf B422} (1994) 382;
Nucl. Phys. {\bf B433} (1995) 390.

\bibitem{BWW} A. Bode, P. Weisz and U. Wolff, Nucl. Phys. {\bf B576}
(2000) 517; erratum-ibid. {\bf B600} (2001) 453.

\bibitem{Gimenez} V. Gimenez, L. Giusti, G. Martinelli and F. Rapuano,
JHEP {\bf 3} (2000) 018.

\bibitem{Reisz} T.~Reisz,
Nucl. Phys. {\bf B318} (1989) 417.

\bibitem{Tarasovetal} O. V. Tarasov, A. A. Vladimirov and A. Zharkov,
Phys. Lett. {\bf B93} (1980) 429.

\bibitem{deWit} B. S. de Wit, Phys. Rev. {\bf 162} (1967) 1195, 1239.

\bibitem{Kluberg} H. Kluberg-Stern and J. B. Zuber, Phys. Rev. {\bf
D12} (1975) 482.

\bibitem{Abbott} L. F. Abbott, Nucl. Phys. {\bf B185} (1981) 189.

\bibitem{ellism} R. K. Ellis and G. Martinelli, Nucl. Phys.
{\bf B235} (1984) 93.

\bibitem{LWbg} M. L\"uscher and P. Weisz, Nucl. Phys. {\bf B452}
(1995) 213.


\bibitem{Ellis} R. K. Ellis, Proceedings of the Workshop: {\em Gauge
Theory on a Lattice}, Eds. C. Zachos et al. (Argonne National
Laboratory, 1984), 191.  

\end{references}
\end{document}